\documentclass[aps,prb,twocolumn,amsmath,amssymb,superscriptaddress,showpacs]{revtex4-1}
\usepackage{graphicx}
\usepackage{dcolumn}
\usepackage{bm}
\usepackage{color}
\newcommand{\ket}[1]{\left| #1 \right>} 
\newcommand{\bra}[1]{\left< #1 \right|} 
\begin{document}
\title{Zero-field and Time-Reserval-Symmetry-Broken Topological Phase Transitions in Graphene}
\author{Marcos R. Guassi}
\author{Ginetom S. Diniz}
\affiliation{Institute of Physics, University of Brasilia, Campus Darcy Ribeiro, DF, 70910-900, Brazil}
\author{Nancy Sandler}
\affiliation{Department of Physics and Astronomy, Ohio University, Athens, Ohio 45701-2979, USA}
\author{Fanyao Qu}
\email{fanyao@unb.br}
\affiliation{Institute of Physics, University of Brasilia, Campus Darcy Ribeiro, DF, 70910-900, Brazil}
\affiliation{Department of Physics, University of Texas at Austin, Austin, Texas 78712, USA}
\date{\today}
\begin{abstract}
We propose a quantum electronic device based on a strained graphene nanoribbon. Mechanical strain, internal exchange field and spin-orbit couplings (SOCs) have been exploited as principle parameters to tune physical properties of the device. We predict a remarkable zero-field topological quantum phase transition between the time-reversal-symmetry broken quantum spin Hall (QSH) and quantum anomalous Hall (QAH) states, which was previously thought to take place only in the presence of finite magnetic field. We illustrate as intrinsic SOC is tuned, how two different helicity edge states located in the opposite edges of the nanoribbon exchange their locations. Our results indicates that the pseudomagnetic field induced by the strain could be coupled to the spin degrees of freedom through the SOC responsible for the stability of a QSH state.  The controllability of this zero-field phase transition with strength and direction of the strain is also demonstrated. Our prediction offers a tempting prospect of strain, electric and magnetic manipulation of the QSH effect.
\end{abstract}

\maketitle
\section{Introduction} 
New classes of matter, such as quantum spin Hall and quantum anomalous Hall states, have been theoretically predicted and experimentally observed in topological insulators \cite{QAHEXP,QAHIT,TIHASAN,QAHNomura}, HgTe-CdTe quantum wells \cite{QSHZHANG1,QSHZHANG2,QSHZHANG3,QSHZHANG4}, graphene \cite{Haldane,kane2005z,PhysRevB.85.115439,young2013tunable} and beyond graphene systems: silicene \cite{QSHYAO,apl4790147}, two-dimensional germanium \cite{QSHYAO,Yandong}, and transition metal dichalcogenides (TMDCs) \cite{ScienceFU,PhysRevLett.113.077201}. Both the QSH and QAH states possess topologically protected edge states at the boundary, where the electron backscattering is forbidden, offering a potential application to electronic devices to transport current without dissipation \cite{kane2005quantum,QAHEXP,QSHZHANG1,ScienceFU}. However, the QSH and QAH states are essentially two very different states of matter. The QSH is characterized by a full insulating gap in the bulk and \emph{helical gapless} edge states where opposite spin counter-propagate at each boundary protected by time-reversal symmetry (TRS) \cite{QSHZHANG1,QSHZHANG2,QSHZHANG3,QSHZHANG4,Haldane,kane2005quantum,kane2005z,QSHYAO}. Whereas in the case of QAH, the helical gapless edge states are replaced by \emph{chiral gapless} edge states where one of the spin channels is suppressed, because of broken TRS \cite{QAHEXP,QAHIT,PhysRevB.84.165453}. Therefore, to realize topological a quantum phase transition (QPT) from the QSH to QAH states, what one needs is to apply a perturbation which can break the TRS \cite{Zhang}. To reach this goal, an external magnetic field is a potential solution. From the application point of view, however, an internal exchange field (EX) which leads to the majority spin band being completely filled while the minority spin band being empty, provides a more attractive alternative way \cite{QAHEXP,PhysRevB.85.115439,diniz2013engineering,oh2013complete}. As known, the strain-induced pseudomagnetic field $B_S$ leads to Landau quantization and edges states that circulate in opposite directions \cite{guinea2009energy,STRAINMAG}. Thus, without breaking TRS, the strain can induce the gap in the bulk and \emph{pseudo-helical} gapless edge states. Therefore, strain, EX and SOC can be used as versatile tools to trigger topological QPTs \cite{Iordanskii,diniz2013engineering}. This motivates us to propose a remarkable way in which SOC strength, uniaxial mechanical strain and EX, instead of external magnetic field, are utilized to realize this QPT in graphene nanoribbons (GNRs).

\begin{figure}[ht]
\begin{center}
\includegraphics[scale=0.70]{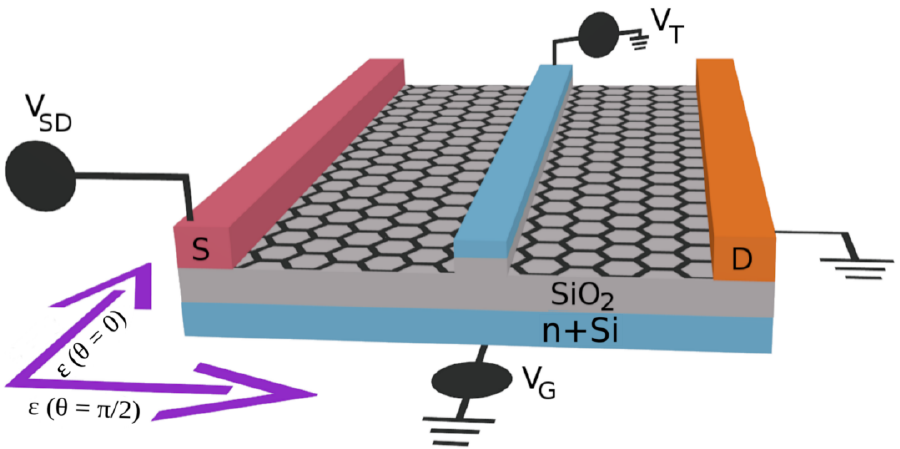}
\caption{Schematic representation of a zigzag GNR (ZGNR) device, deposited on $SiO_2$. Electrical current is controlled between S and D using ($V_{SD}$) bias. The uniaxial strain is applied along either directions indicated by the arrows. The Rashba SOC strength could be tuned by top ($V_T$) and bottom ($V_G$)voltages.}
\label{fig1}
\end{center}
\end{figure}

\section{Theoretical Model}
The system is schematically illustrated in Fig. \ref{fig1} and described by the following tight-binding Hamiltonian,
\begin{align}
\label{H0}
H = & -\sum_{\left< i,j \right>} t_{i,j}c_{i}^{\dagger} c_{j}
     + \frac{2i}{\sqrt{3}} \lambda_{so} \sum_{\left< \left< i,j \right> \right>} c_{i}^{\dagger} \vec{\gamma}\cdot(\vec{d}_{kj} \times \vec{d}_{ik}) c_{j}\\
	& +i \lambda_R \sum_{\left< i,j \right>}  c_{i}^{\dagger } \hat{e}_{z}\cdot( \vec{\gamma} \times \vec{d}_{ij})  c_{j} \nonumber
	 + M \sum_{i} c_{i}^{\dagger} \gamma_{z} c_{i}.
\end{align}Here, $c_i^{\dagger}$ ($c_{i}$) is the $\pi$-orbital creation (annihilation) operator for an electron on site $i$, where the spin index on the electron operators has been suppressed, $d_{ij}$ is a lattice vector pointing from
site $j$ to site $i$, $\vec{\gamma}$ is a vector whose components are the Pauli matrices and $\hat{e}_{z}$ is a unit vector along the $z$-direction. $< >$ ($<<  >>$) runs over all the nearest (next-nearest) neighbor sites. The first term is the nearest neighbors hopping term on the honeycomb lattice with hopping amplitude $t_{i,j}$. The second term denotes the intrinsic SOC with coupling strength $\lambda_{so}$ \cite{kane2005quantum}, predicted to be rather small in pristine carbon structures due to the low atomic number of carbon atoms. However, recent experiments have demonstrated that it can be enhanced up to three orders of magnitude, which is about 17 meV by the proximity effect to TDMCs \cite{NatCom}, with no drastic modification of the structure of the graphene, or by adding adatoms \cite{NatCom2}, such as covalently bonding hydrogen atoms to the graphene lattice \cite{NatPhys}. The third one represents the Rashba SOC with strength $\lambda_{R}$ \cite{Zarea}, whose values ranging from 13-225 meV have been experimentally reported on different setups \cite{PhysRevLett.101.157601,PhysRevLett.100.107602,PhysRevLett.102.057602}. The last term corresponds to the EX with strength $M$, that might be achieved by magnetic atom doping in the graphene lattice \cite{PhysRevLett.110.136804} or due to proximity effect by coupling the graphene to ferromagnetic insulators \cite{Swartz}, such as a thin film $BiFeO_3$ for which an estimate of the exchange field predicts a value of 70 meV \cite{PhysRevLett.112.116404}.

The uniaxial strain may be induced either by an external stress applied to the GNR in a particular direction \cite{NanoBao,nl403679f} or by a substrate due to deposition of graphene on top of other materials \cite{apl3463460,PhysRevB.79.205433,nn800031m}. The strain modified distances between carbon atoms are described by $\vec{d}^s_i=(I+\epsilon)\vec{d}_i$, with $\vec{d}_{i}$ (i=1, 2, 3) the unstrained vectors for nearest-neighbors, $I$ is the identity matrix and $\epsilon$ is the strain tensor defined as \cite{pereira2009tight},
\begin{eqnarray}
\epsilon=
\varepsilon\left(\begin{array}{cc}
\cos^{2}\theta -\nu\sin^{2}\theta & (1+\nu)\cos\theta\sin\theta\\
(1+\nu)\cos\theta\sin\theta & \sin^{2}\theta -\nu\cos^{2}\theta\\
\end{array}\right)
\end{eqnarray}
where $\nu=0.165$ is the Poisson's ratio value known for graphite \cite{pereira2009tight}, $\theta$ is the direction of strain and $\varepsilon$ is the strain modulus. Then, the hopping term is affected by the strain through $t_{i,j}=t_i=t e^{-3.37(d^s_i/a-1)}$, with $t=2.7$eV \cite{pereira2009tight} being the unstrained hopping parameter and $a$ is the C-C distance. We define the direction as $\theta = 0$ when it is parallel to the zigzag chain and $\theta = \pi/2$, when it is along armchair direction, as shown in Fig. \ref{fig1}.
Before proceeding to the GNR cases, we can make an analysis in the \textit{bulk} graphene by performing a Fourier transformation of Eq. \ref{H0}, resulting in a $4\times4$ Hamiltonian matrix $H(\vec{k})$. In the low energy limit, we expand the momentum at the vicinity of the Dirac points, $\vec{k}= \eta \textbf{K} + \vec{q}$, where $\textbf{K}=(K_{x},K_{y})$ are the strain-shifted Dirac points obtained by using the condition $\textbf{K}\cdot\left(\vec{d}_{1}^{s}-\vec{d}_{2}^{s}\right)=\cos^{-1}(t_{3}^{2}-t_{1}^{2}-t_{2}^{2}/2t_{1}t_{2})$, with $\eta$=$\pm$1 related to the two valleys \cite{PhysRevB.81.035411} and $\vec{q}=(q_x,q_y)$ is a small crystal momentum around $\eta \textbf{K}$. Notice that the permutation of $d_{i=1,2,3}^{s}$ and $t_{i=1,2,3}$ also satisfies the previous relation to obtain the strain-shifted Dirac points. We can write the full Hamiltonian in the basis of $\left\{\Psi_{A}(\eta K,\uparrow), \Psi_{A}(\eta K,\downarrow), \Psi_{B}(\eta K,\uparrow), \Psi_{B}(\eta K,\downarrow)\right\}^{\dagger}$ as
\begin{eqnarray}
H(\vec{q})=\left(\begin{array}{cc}
t_{so} + MS_{z}                & f + t_{R}\\
f^{*} + t_{R}^{*} &        -t_{so}+ MS_{z}\\
\end{array}\right),
\end{eqnarray}
where $f$, $t_{so}$ and $t_{R}$ are the strain dressed hopping, intrinsic- and Rashba- SOCs, respectively given by \cite{diniz2013engineering}
\begin{widetext}
\begin{eqnarray}
f&=&-\{t_{1}[1-i(1+\epsilon_{22})q_{y} -i\epsilon_{12}q_{x}] e^{-i2\eta \theta_{1}} \\\nonumber
     &+&t_{2}[1+i/2(\epsilon_{12}+\sqrt{3}(1+\epsilon_{11}))q_{x} + i/2(\sqrt{3}\epsilon_{21}+1+\epsilon_{22}))q_{y}] e^{i\eta\theta_{3}^{+}}\\\nonumber
     &+&t_{3}[1+i/2(\epsilon_{12}-\sqrt{3}(1+\epsilon_{11}))q_{x} - i/2(\sqrt{3}\epsilon_{21}-1-\epsilon_{22}))q_{y}] e^{i\eta\theta_{3}^{-}}\}\textbf{1}_{s},\\\nonumber
\quad\\\nonumber
t_{so}&=&det\left[I+\epsilon\right]\eta\lambda_{so}\{2\sin{(2\theta_{2})}-4\sin{(\theta_{2})}\cos{(3\theta_{1})}\}S_{z},\\\nonumber
\quad\\\nonumber
t_{R}&=&\lambda_{R}\{[-i(1+\epsilon_{22})e^{-2i\eta\theta_{1}}-(\sqrt{3}\eta\epsilon_{21}\sin{\theta_{2}}-i(1+\epsilon_{22})\cos{\theta_{2}})e^{i\eta\theta_{1}}]S_{x} \\\nonumber
 &+&[i\epsilon_{12}e^{-2i\eta\theta_{1}} + (\sqrt{3}\eta(1+\epsilon_{11})\sin{\theta_{2}}-i\epsilon_{12}\cos{\theta_{2}})e^{i\eta\theta_{1}}]S_{y}\}.\\\nonumber
\end{eqnarray}
\end{widetext}
Here, $\epsilon_{ij}$ are the matrix elements of the strain tensor $\epsilon$, $\textbf{1}_{s}$ is the identity matrix, $S_{z}$ is the Pauli spin matrix in the real spin subspace, $det\left[I+\epsilon\right]=(1-\epsilon_{11})(1-\epsilon_{22})-\epsilon_{21}\epsilon_{12}$, $\theta_{1}=1/2\left[\epsilon_{12}K_{x} + (1+\epsilon_{22})K_{y}\right]$, $\theta_{2}=\sqrt{3}/2\left[(1+\epsilon_{11})K_{x} + \epsilon_{21}K_{y}\right]$ and $\theta_{3}^{\pm}=1/2\left[(\epsilon_{12}\pm\sqrt{3}(1+\epsilon_{11}))K_{x}\pm (\sqrt{3}\epsilon_{21}\pm1\pm\epsilon_{22})K_{y}\right]$. In a special case in which there is vanishing strain, i.e., $\varepsilon\rightarrow0$ and $(K_{x},K_{y})=(\eta 4\pi/3\sqrt{3},0)$, with $a$ set as unity for simplicity, our expression of the Hamiltonian reduces to the well-known and expected result, as obtained in Ref. 11.
The band gap at the shifted Dirac points, is then given by:
\begin{equation}
\Delta_{K}=\Delta_{K^{\prime}}=\vert-2\phi_{so}+\sqrt{M^{2} + \vert\phi_{R1}\vert^{2}}+\sqrt{M^{2} + \vert\phi_{R2}\vert^{2}}\vert,
\end{equation}
where we have defined: $\phi_{so}=det\left[I+\epsilon\right]\lambda_{so}\{2\sin{(2\theta_{2})}-4\sin{(\theta_{2})}\cos{(3\theta_{1})}\}$, $\phi_{R1}=R_{1}-iR_{2}$ and $\phi_{R2}=R_{1}^{*}-iR_{2}^{*}$, with $R_{1}=\lambda_{R}[-i(1+\epsilon_{22})e^{-2i\eta\theta_{1}}-(\sqrt{3}\eta\epsilon_{21}\sin{\theta_{2}}-i(1+\epsilon_{22})\cos{\theta_{2}})e^{i\eta\theta_{1}}]$ and $R_{2}=\lambda_{R}[i\epsilon_{12}e^{-2i\eta\theta_{1}} + (\sqrt{3}\eta(1+\epsilon_{11})\sin{\theta_{2}}-i\epsilon_{12}\cos{\theta_{2}})e^{i\eta\theta_{1}}]$. For systems with mirror symmetry, $\lambda_{R}$ becomes zero. Then we obtain $\Delta_{K}=\Delta_{K^{\prime}}=2\vert -\phi_{so} + M \vert$, from which the critical exchange field $M_{C}^{(s)}=\phi_{so}$ can be straightforwardly derived.

To identify the topological properties of the Dirac gap and study the origin of QAH, we have also calculated the Berry curvature $\Omega_{xy}^{n}(k_{x},k_{y})$ of the n$th$ bands using the Kubo formula
\begin{equation}
\Omega_{xy}^{n}(k_{x},k_{y})=-\sum_{n^{\prime}\neq n}\dfrac{2Im\langle \Psi_{nk}\vert v_{x}\vert \Psi_{n^{\prime}k}\rangle\langle \Psi_{n^{\prime}k}\vert v_{y}\vert \Psi_{nk}\rangle}{(\omega_{n^{\prime}}-\omega_{n})^{2}},
\label{Berry}
\end{equation}
where $\omega_{n}=E_{n}/\hbar$ with $E_{n}$ the energy eigenvalue of the n$th$ band and $v_{x(y)}={\hbar}^{-1}\partial H/\partial k_{x(y)}$ is the Fermi velocity operator. The Chern number $\mathcal{C}$ can be calculated by \cite{PhysRevLett.49.405}
\begin{equation}
\mathcal{C}=\dfrac{1}{2\pi}\sum_{n}\int_{BZ}d^{2}k \Omega_{xy}^{n},
\label{H4}
\end{equation}
where the summation is taken over the occupied states below the Fermi level and the integration is done over the first Brillouin zone.
As the Berry curvatures are highly peaked around the Dirac points $\textbf{K}$ and $\textbf{K}^{\prime}$ \cite{nl300610w}, then a low energy approximation can be used in the calculation of the Chern number \cite{PhysRevB.82.161414,diniz2013engineering}. In the low energy, we calculate the Chern number using the following equation,
\begin{equation}
\mathcal{C}=\dfrac{1}{2\pi}\sum_{K,K^{\prime}}\sum_{n=1,2}\int_{-\infty}^{{\infty}}dq_{x}dq_{y} \Omega^{n}_{xy}(q_{x},q_{y}).
\label{H4}
\end{equation}
In the above integral, a momentum cutoff is set around each valley for which the Chern number converges.
\section{Results}
\begin{figure}
\includegraphics[scale=0.46]{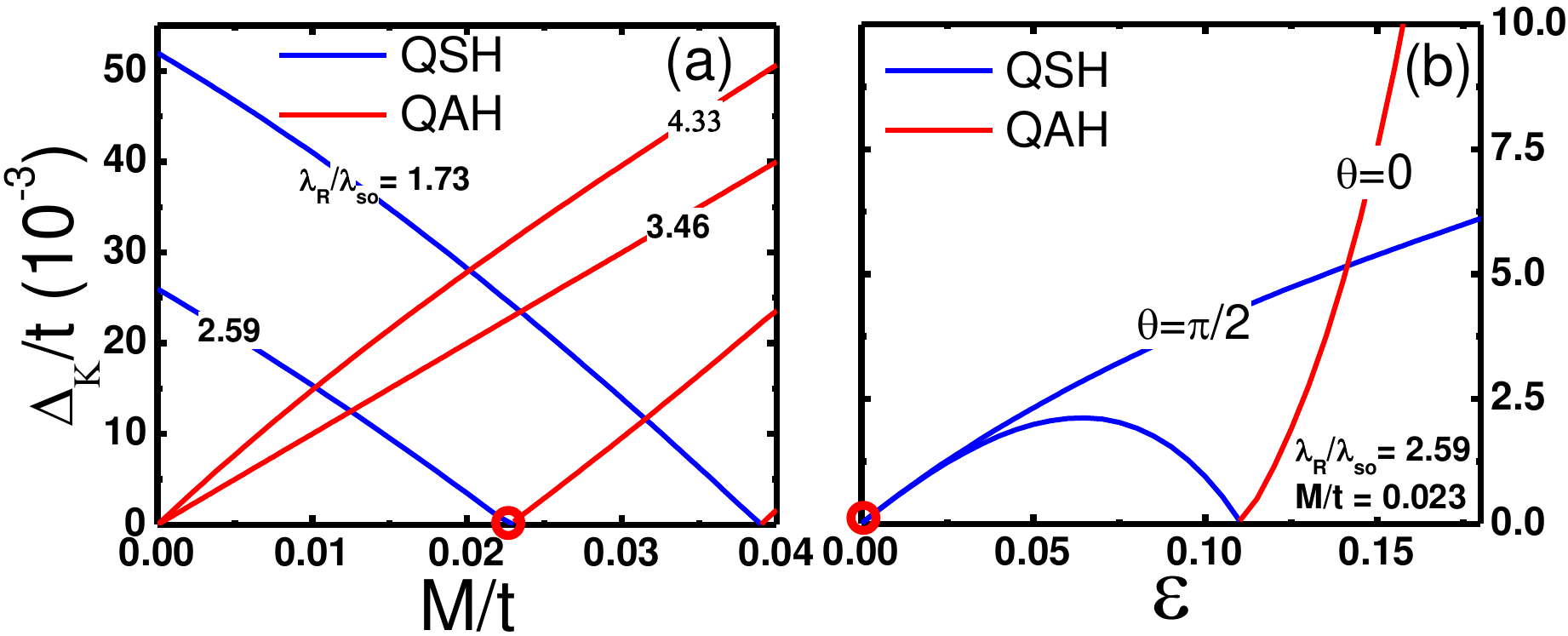}
\caption{(a) Energy band gap $\Delta_{K}$ ($\Delta_{K}=\Delta_{K^{\prime}}$) as a function of exchange field M for graphene with four different values of Rashba SOC strength $\lambda_R/\lambda_{so}$ for $\varepsilon$=0 and (b) $\Delta_{K}$ versus uniaxial strain $\varepsilon$ applied along $\theta = 0$ and $\theta = \pi/2$ for a strained graphene with $\lambda_R/\lambda_{so} = 2.59$ and $M/t=0.023$.  QSH (blue lines) and QAH (red lines) phases are characterized by Chern numbers $\mathcal{C}$=0 and 2, respectively. The circles indicate the critical point at which the phase transition occurs.}
\label{fig2}
\end{figure}

Fig. \ref{fig2} (a) plots energy band-gap $\Delta$ between the conduction and valence bands of graphene as a function of EX for different values of $\lambda_R/\lambda_{so}$. In the regime of $\lambda_{so}$ comparable to $\lambda_R$ ($\lambda_R< 2 \sqrt{3} \lambda_{so}$) and EX, the band gaps at either $K$ or $K^{\prime}$ for unstrained bulk graphene can be well described by $\Delta_K=\Delta_{K'}=\vert\sqrt{M^2+9\lambda_R^2} + M - 6\sqrt{3}\lambda_{so}\vert$ \cite{PhysRevB.85.115439}. Hence, for a given $\lambda_R$, the gap first decreases with increasing EX and then closes when EX reaches a critical value, $M_C = \frac{\sqrt{3} \lambda_{so}}{4} \left[ 12 - \left( \frac{\lambda{R}}{\lambda_{so}} \right)^2 \right]$, as shown by circles in Fig. \ref{fig2} (a)-(b). After that, as EX is further increased, the gap reopens, accompanied by a change of the Chern number $\mathcal{C}$ as demonstrated by the color-change of the correspondent line \cite{diniz2013engineering,PhysRevB.85.115439,PhysRevLett.112.116404,PhysRevB.84.165453,TIHASAN}. Therefore, a QPT between QSH to QAH occurs at a critical exchange field $M_C$. After thoroughly understanding the fate of the TRS-broken QSH phase in zero-strain graphene, we move our attention to strained graphene. We found that the band gap as well as the critical point are strongly affected by the applied mechanical strain field. For instance, in the absence of the Rashba SOC, the critical value $M_C^{s}$ of strained graphene is given by $M_C^{(s)} = \phi_{so}$, where $\phi_{so}$ is the strain-dressed intrinsic SOC strength. In comparison with $M_C$, we notice that the critical point is shifted by strain, as shown in Fig. \ref{fig2} (b). Notice that at $M=M_C$, highlighted by the red circle, is no longer the value for the critical point for strained graphene, because no phase transition occurs at this point. Although the strain widens the bulk gap monotonically in the case of strain along the direction $\theta = \pi/2$, the closing and reopening phenomenon; and consequently a phase transition between QSH and QAH; is found for the strain applied along $\theta = 0$ direction.

\begin{figure}
\begin{center}
\includegraphics[scale=1.1]{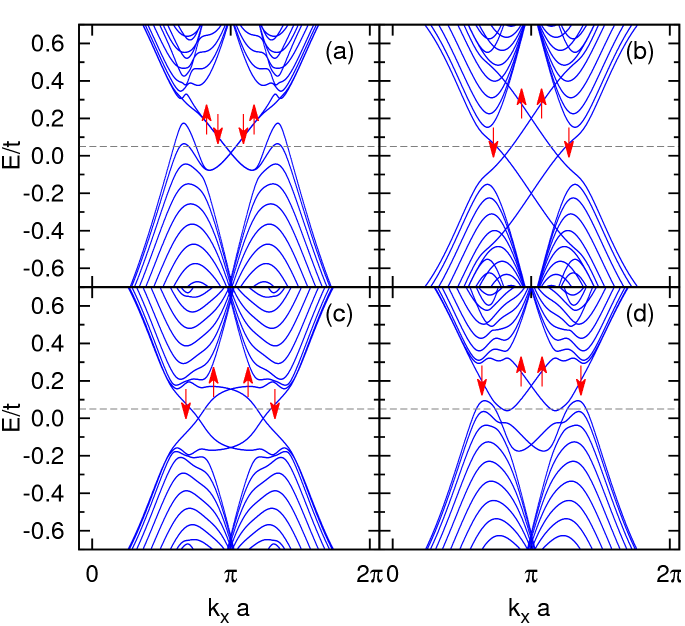}
\caption{Band structure of ZGNR with intrinsic- and Rashba- SOC terms (a), intrinsic SOC and EX (b), Rashba SOC and EX (c), and intrinsic- and Rashba- SOCs and EX (d). The Fermi level is assumed to be above zero, as indicated by the dashed horizontal line, and thus has four intersections with the conduction bands. This gives rise to four edge currents on the ribbon edges. The following parameters are used: (a) $\lambda_{so} = 0.06t$, $\lambda_R = 0.20t$; (b) $\lambda_{so} = 0.06t$, $M=0.20t$; (c) $\lambda_R = 0.20t$, $M=0.20t$; (d) $\lambda_{so} = 0.06t$, $\lambda_R = 0.20t$ and $M=0.20t$ for the ZGNR with width $W=48$. The arrows represent the major components of spin.}
\label{fig3}
\end{center}
\end{figure}

If the mirror symmetry about the graphene-plane is preserved, then the intrinsic SOC which opens gaps around Dirac points is the only allowed spin dependent term in the Hamiltonian. Otherwise, if the mirror symmetry is broken, then a Rashba term is allowed, which mixes spin-up and spin-down states around the band crossing points. Besides, Rashba SOC pushes the valence band up and the conduction band down, reducing the bulk gap. Following Ref. 19, we present our results for the ZGNR in Fig. \ref{fig3}, that shows the effects of intrinsic- and Rashba- SOCs and EX upon the band structure of the ZGNR. Notice in Fig. \ref{fig3} (a) that the interplay between intrinsic- and Rashba- SOCs, partially lifts the degeneracies of both bulk- and edge- state, breaks particle-hole symmetry and pushes the valence band up. In turn, the presence of the EX breaks the TRS and lifts the Kramer's degeneracy of electron spin, pushing the spin-up (spin-down) bands upward (downward), as shown in Fig. \ref{fig3} (b). In strong contrast with Fig. \ref{fig3} (b), the presence of Rashba SOC and EX induces coupling between edge and bulk states which significantly modifies the group velocity of edge states, as shown in Fig. \ref{fig3} (c). The combined effects of intrinsic, Rashba SOCs and EX are shown in Fig. \ref{fig3} (d), which are in agreement with results reported in Ref. 19 (see for instance Fig 2). Notice that the Fermi level enters into the valence band and the energies of some edge modes are smaller than the valence band maximum.

\begin{figure}
\begin{center}
\includegraphics[scale=1.1]{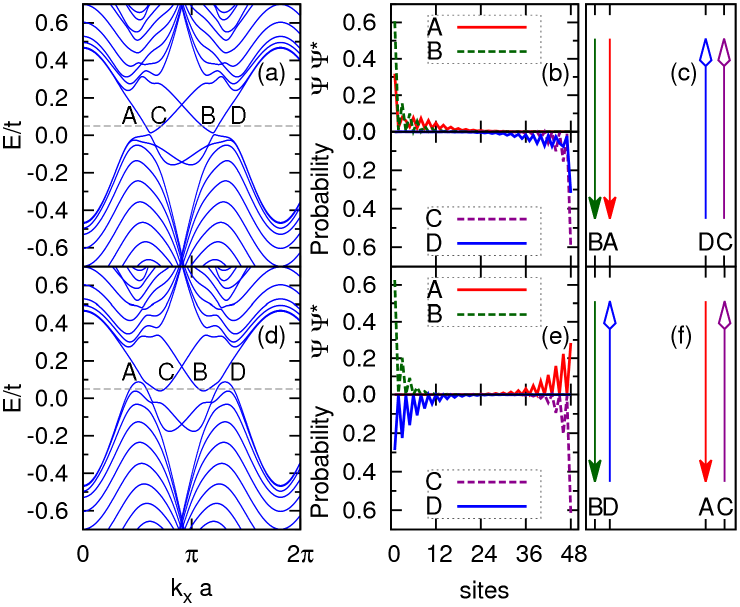}
\caption{Energy spectrum of ZGNR with $W=48$, $\lambda_R = 0.20t$, $M = 0.20t$, $\varepsilon = 0.10$ and $\theta = 0$, for (a) $\lambda_{so}= 0.035t$  and (d) $\lambda_{so}  = 0.055t$, respectively. The Fermi level $E = 0.05t$ corresponds to four different edge states, as indicated by A, B, C, and D. The corresponding probability distributions $|\psi|^2$ across the width of the ribbon, and diagrams of charge current distributions are shown in the middle (b)-(e) and right panels (c)-(f), respectively. The arrows indicate the current flux.}
\label{fig4}
\end{center}
\end{figure}
The intrinsic SOC can be strongly enhanced by impurity (adatom) coverage on the surface of graphene, that produces strong lattice distortions \cite{neto2009impurity}. In this context, one may ask how the quantum phase transition in a graphene ribbon changes as the intrinsic SOC is tuned. Following the discussion of Ref. 19, the effects of strain fields are shown in Fig. \ref{fig4} (with a similar representation to the one introduced in Ref. 19) with parameters W=48, $\lambda_R = 0.20t$, $M = 0.20t$ and uniaxial strain $\varepsilon = 0.10$ along $\theta = 0$. The left panel of Fig. \ref{fig4} shows the effects of intrinsic SOC on the energy spectrum of a ZGNR. The Fermi level is set at $E_F = 0.05 t$. The corresponding edge state probability distributions across the width of the nanoribbon, for each of the four edge states indicated by A, B, C and D  are shown in the middle panel. Schematic diagrams of charge current distributions on the edges of ZGNR are illustrated in the right panel. To determine the edge current direction, $I = -|e| v_x$ (indicated by the arrow), the electron group velocity $v_x=\partial E(k)/\partial k_x$ has been calculated \cite{PhysRevB.84.165453}. In the case of weak intrinsic SOC, at the ribbon boundaries, the edge states pair A and D would form a single handed loop (in the sense that the turning point is at infinity along the ribbon length), meanwhile there is the formation of another loop with opposite handedness, which is formed by the edge states pair B and C. Both edge states A and B, consequently $I_A$ and $I_B$, are located at the same edge, as indicated in Fig. \ref{fig4} (c). Thus the chirality of the current loop due to the A and D edge states would produce a Chern number of $(\mathcal{C}_1=\pm 1)$ which is the same as that of current loop owing to B and C edge states. Since the system is akin to two integer quantum Hall subsystems, its Chern number $\mathcal{C}$ is equal to $(\mathcal{C}_1= +1) \oplus (\mathcal{C}_2 = +1)$, i.e., $\mathcal{C}= (+1) + (+1) = 2$ or $(\mathcal{C}_1= -1) \oplus (\mathcal{C}_2= -1)$, with $\mathcal{C}= (-1) + (-1) = -2$. Therefore, the ZGNR with a weak intrinsic SOC is in the QAH phase.  For a ZGNR with strong intrinsic SOC, however, one can notice that the edge states pair A and C are located on the same edge, while the B and D edge states are in the opposite edge, as shown in Fig. \ref{fig4} (f). Due to handedness of the current loop of edge states A and D, the Chern number would give a contribution of $(\mathcal{C}_1=-1)$, and the pair B and C, which has an opposite handedness, produces a Chern number of $(\mathcal{C}_2=+1)$. Since the ZGNR is composed of these two integer quantum Hall subsystems, its Chern number $(\mathcal{C})$ is obtained by $(\mathcal{C}_1= +1) \oplus (\mathcal{C}_2 = -1)$, i.e., $\mathcal{C} = (+1) + (-1) = 0$. Therefore, the GNR is in the TRS broken QSH phase.

\begin{figure}
\includegraphics[scale=1.0]{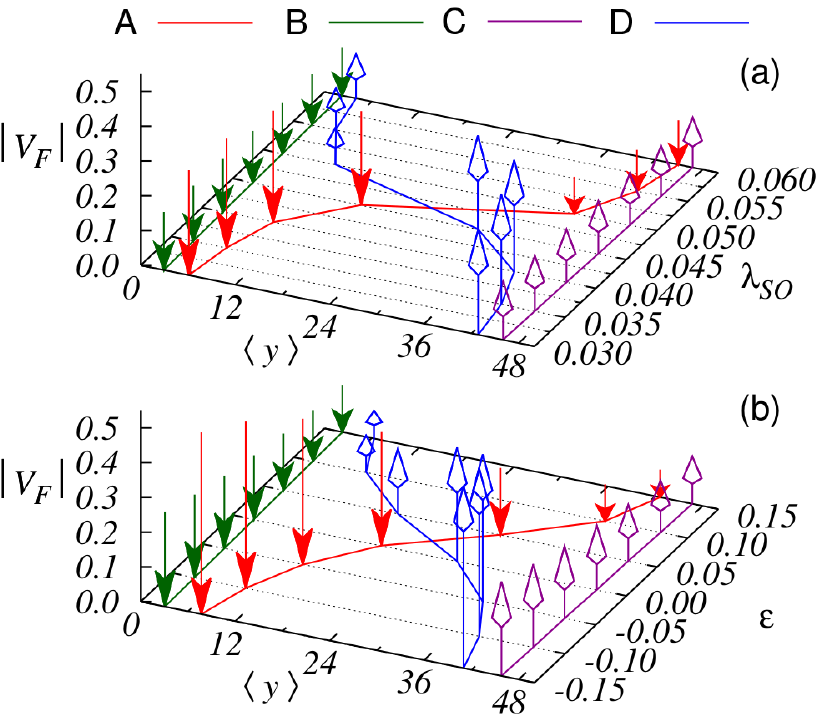}
\caption{(a) Average values $\left< y \right>$ of edge states versus $\lambda_{so}$ in ZGNR, subjected to a strain with $\varepsilon = 0.10$ and $\theta = 0$. (b) $\left< y \right>$ as a function of strain with $\theta = 0$ for $\lambda_{so} = 0.05t$. $W=48$, $\lambda_R = 0.20t$ and $M=0.20t$ are used in the computations. Vertical axis is the Fermi velocity $V_F$ modulus. The arrows point in the directions of band velocities and their lengths present the magnitudes of $V_{F}$.}
\label{fig5}
\end{figure}

To understand the QPT and show intuitively how it takes place, we follow Ref. 19 and introduce the average value of the position $\left< y \right>_{n}$, as a parameter to label the angular momentum of the current. It is defined as: $\left< y \right>_{n}= \sum_i y_i |\varphi_{n} (y_i)|^2$, where $n$ represents the edge states at the Fermi level and $i$ is the site index along the width of ribbon. We chose the origin of $y$-axis at the lower boundary of the ribbon.  Fig. \ref{fig5} (a) shows the average values $\left< y \right>_n$ of edge states as a function of $\lambda_{so}$ in the ribbon with the width $W=48$, $\lambda_R = 0.20t$, $M=0.20t$, $\varepsilon = 0.10$ and $\theta = 0$, where $n= A, B, C$ and $D$, respectively. The direction and magnitude of a group velocity are indicated by the direction and length of an arrow, respectively. When the intrinsic SOC is vanishing, the Rashba SOC and EX are dominant, A and B are on the same boundary of the ribbon, and thus both $\left< y \right>_A$ and $\left< y \right>_B$  $\rightarrow 0$. So do C and D, but are localized at the other edge of the ribbon, thus $\left< y \right>_C$ and $\left< y \right>_D$  $\rightarrow W$. The system is in the QAH phase. When the $\lambda_{so}$ increases, however, three different topological phases are found. In the regime of small $\lambda_{so}$ ($0.03t <\lambda_{so}< 0.04t$), the positions of the edge states are only very slightly shifted. With increasing $\lambda_{so}$, the states A and D become delocalized, swiftly moving to the center of the ribbon from different boundaries owing to the edge- and bulk- states coupling. In the regime of large $\lambda_{so}$ ($\lambda_{so} > 0.05$), the locations of state A and D have been exchanged. Since the group velocity of state A is opposite to D, the exchange of their locations results in a change of chirality. Therefore, the system is in the QSH phase. It is worthy to point out that owing to the finite-size (finite-width) effect, the edge states are not exactly localized at the two boundaries. Remarkably, a similar behavior is also presented in Fig. \ref{fig5} (b) in which $\left< y \right>$ versus strain is plotted. At first glance, it seems to be hard to understand this exotic behavior. But, recalling the discussion of phase transition in bulk graphene, one can logically speculate that this is a manifestation of strain induced QPT between QSH and QAH states in the ZGNR. This strain induced QSH state shares many emergent properties similar to the usual zero-strain QSH effect. We notice that with realistic values for uniaxial strain the critical value for the spin-orbit coupling is reduced by a factor between 10-20\%. Thus, the combination of strain and appropriate substrates, show a promising direction to realize the phase transition in current settings.

\begin{figure}
\includegraphics[scale=1.0]{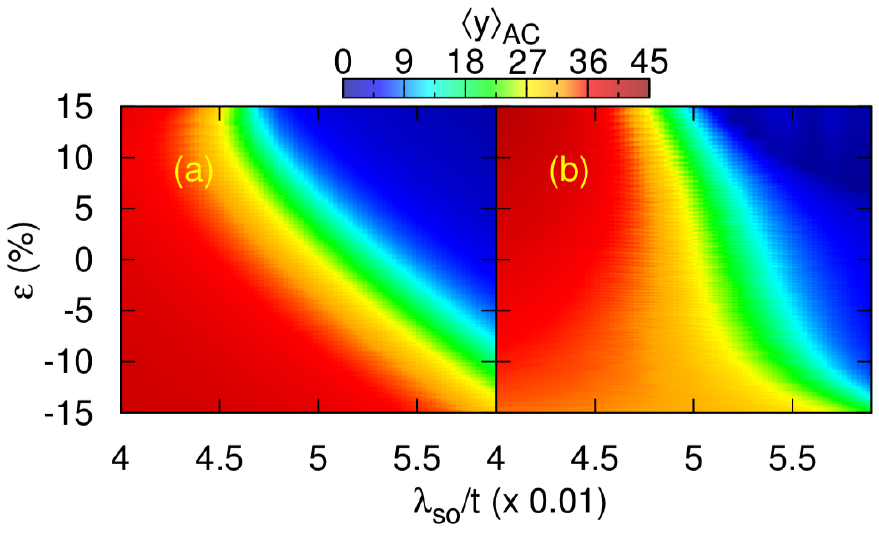}
\caption{Phase diagrams (strain vs intrinsic SOC) of a ribbon with $W=48$, $\lambda_R = 0.20t$, $M = 0.20t$, characterized by a difference in the average value of position  between mode A and C, defined as $\left< y \right>_{AC}$= $\left< y \right>_C$ - $\left< y \right>_A$, for $\theta=0$ (a) and $\theta=\pi/2$ (b), respectively.}
\label{fig6}
\end{figure}

To seek the controllable topological QPTs induced either by strain, EX; or intrinsic SOC, or any of their combinations, the phase diagrams in which the phase is characterized by the difference in the average value of position $\left< y \right>_C$ and $\left< y \right>_A$, defined as $\left< y \right>_{AC}$ = $\left< y \right>_C$ - $\left< y \right>_A$, are constructed, as shown in Fig. \ref{fig6}. Fig. \ref{fig6} (a) and (b) plot the phase diagrams of $\varepsilon$ versus $\lambda_{so}$ for $\theta = 0$ and $\theta = \pi/2$, respectively. It is trivial to notice that if $\left< y \right>_{AC} \cong 0$, the edge states A and C are localized at the same boundary, corresponding to a QSH phase, as indicated by blue. Otherwise, if $\left< y \right>_{AC} \cong W$, the system is in the QAH phase, as marked by red. The other values of $\left< y \right>_{AC}$ correspond to delocalized state A. Notice that both strength and direction of the strain change considerably the phase diagram. In the regime of small intrinsic SOC, the GNR lies in the QAH state. The critical $\lambda_{so}^{c}$ at which topological QPT occurs depends strongly on both the strength and direction of the strain. The larger the strain, the smaller the  $\lambda_{so}^{c}$ is required to reach the QSH state. In addiction, the strain drives the GNR from the QAH into QSH states for a given $\lambda_{so}^{c}$. It is also noted that in the case of $\theta = \pi/2$, when the $\lambda_{so}^{c}$ changes in the boundary between the QSH and QAH states, the correspondent critical value of $\varepsilon$ varies faster than that for $\theta = 0$.

The underlying physics of the strain tuned phase diagram is as follows. It is well established that uniaxial mechanical strain does not break the sublattice symmetry, but rather deforms the Brillouin zone, such as, the Dirac cones located in graphene at points $K$ ($K^\prime$) being shifted in the opposite directions \cite{pereira2009tight,PhysRevB.88.085430}. This is reminiscent of the effect of pseudomagnetic field $B_S$ induced by the strain on charge carriers, i.e., accumulating charge in place where the $B_S$ is maximum. Because the $B_S$ does not break TRS, the strain will not have any direct effect on the spin degrees of freedom of the electrons, even though it couples with sublattice pseudospin. Therefore, at first glance, it seems that the strain only induces a renormalization of the energy scales. Actually, \emph{this is not true} for graphene with SOC. Since SOC couples the spin and the momentum degrees of freedom of the carriers, the $B_S$ could affect real spin of an electron through the SOC. Therefore, a strong pseudomagnetic field should lead to Landau quantization and a QSH state due to opposite signs of $B_S$ for electrons in valleys $K$ ($K^\prime$). In this context, the strain enhances the carrier localization and pushes the edge states much closer to the boundaries of the ribbon. Hence, the QSH state could be stabilized by the strain. Finally, it is worthwhile to argue that since inter-valley scattering requires a large momentum transfer \cite{PhysRevB.85.245418}, it is strongly suppressed in wide ZGNRs in which we are interested.
%
\section{Conclusion}
In summary, a zero-field topological QPT between QSH and QAH states in GNRs is reported in the presence of internal EX, uniaxial strain, intrinsic and Rashba SOCs. Both strength and direction of the strain can be exploited to tune the $\lambda_{so}^{c}$ at which the phase transition takes place. The pseudomagnetic field induced by the strain couples the spin degrees of freedom through SOC, enhances the carrier localization in edge states, stabilizes and even leads to formation of a QSH state. Rashba-SOC and EX, on the other hand, break inversion- and TRS of the graphene, respectively. In the regime of small SOC and EX, they only induce an instability of the QSH state. The large Rashba-SOC or EX, however, can even lead the QSH state to be destroyed, producing the QAH states. Our results offer a tempting prospect of strain, electric and magnetic manipulation of the QSH effect, with potential application in topological quantum devices within the context of dissipationless electronics.

\section{Acknowledgments}
We thank fruitful discussions with M. Ezawa and Z. Qiao. We acknowledge financial support received from CAPES, FAP-DF and CNPq (MRG, GSD and F. Qu) and NSF MWN/CIAM grant DMR-1108285 (NS).
\appendix
\section{Equation of motion of an electron in strained graphene}
\subsection{Strained graphene}
\begin{figure}
\includegraphics[scale=0.43]{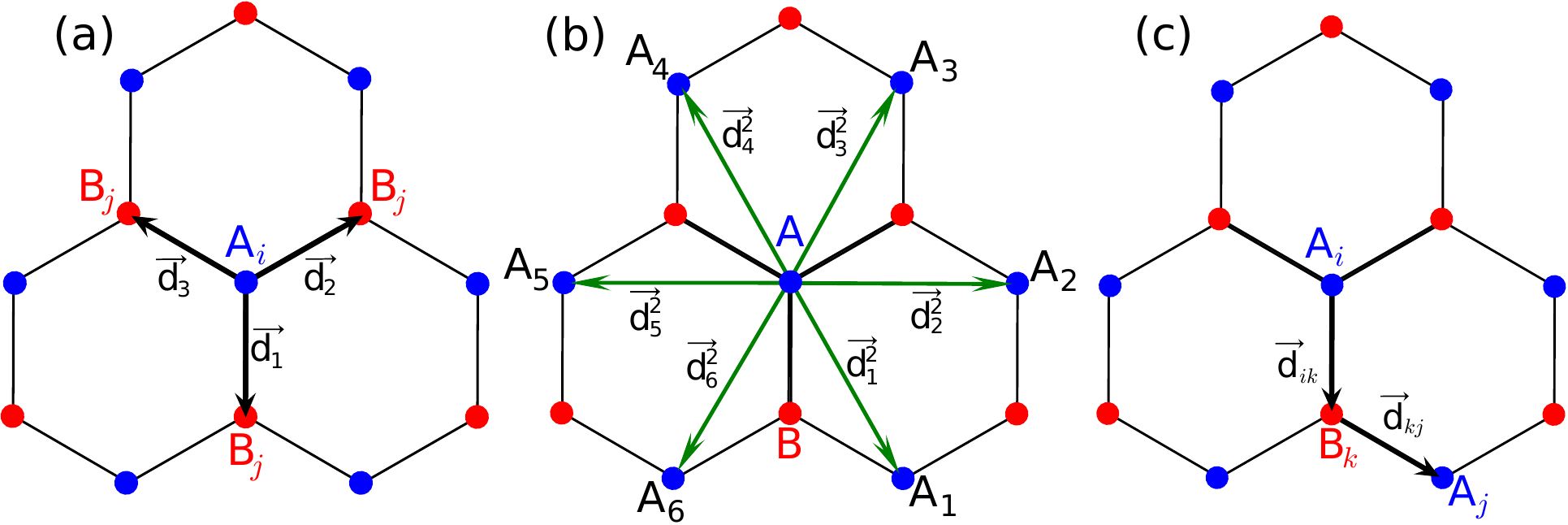}
\caption{Schematic diagrams of the nearest (a) and next nearest neighbours (b), and interatomic distance vectors (c) in a graphene.}
\label{fig7}
\end{figure}

In the pristine (unstrained) graphene, as usual, the Hamiltonian which describes the hopping of an electron in a site $A_i$ to its nearest-neighbors in $B_j$ with probability $t_{i,j}$ is given by,
\begin{align}
	H_0 = - \sum_{\left<i,j \right>} t_{i,j} \left( \ket{A_{i}} \bra{B_{j}} + \ket{B_{j}} \bra{A_{i}} \right)
	\label{hamiltonianoGrafeno}
\end{align}
where the sum is made about the three nearest neighbors. For convenience, the site energy which is on Fermi level is set to zero. The vectors connecting a type A to type B sites are defined by $\vec{d}_1 = - a \hat{y}, \vec{d}_2 =  \frac{\sqrt{3}}{2} a \left( \hat{x} + \frac{1}{\sqrt{3}} \hat{y} \right)$ and $\vec{d}_3 =  \frac{\sqrt{3}}{2} a \left( - \hat{x} + \frac{1}{\sqrt{3}} \hat{y} \right) $, as shown in Fig. \ref{fig7}(a). We expand single particle wave functions as follows
\begin{align}
\ket{\Psi} = \frac{1}{\sqrt{N}} \sum_{n} e^{i \vec{k} \cdot \vec{R}_n} \left[ \Psi_{A}(\vec{k}) \ket{A_{n}} + \Psi_{B}(\vec{k}) \ket{B_{n}} \right].
\label{vetOnda}
\end{align}
where $\vec{R}_n$ is the position of a site-$n$, $\vec{k}$ is electron momentum, $\Psi_{A}(\vec{k})$ and $\Psi_{B}(\vec{k})$ are coefficients. Utilizing $\vec{R}_{n'} - \vec{R}_n = \vec{d}_l$, Schr\"odinger's equation can be cast into two coupled equations:
\begin{align}
E \Psi_{A}(\vec{k}) = f \cdot \Psi_{B}(\vec{k}) \notag
\end{align}
and
\begin{align}
E \Psi_{B}(\vec{k}) = f^{*} \cdot \Psi_{A}(\vec{k}) ,
\label{final_esq_1}
\end{align}
where the geometric form factor is given by
\begin{align} 	
f = - \sum_l^3  t_l \ e^{i \vec{k} \cdot \vec{d}_l} = - t \left[ e^{-i k_y a} + 2 cos \left( \frac{k_x a \sqrt{3}}{2} \right) e^{i k_y a/2} \right] . \notag
\end{align}
Based on above calculation, one can straight forwardly derive the matrix form of the Schr\"odinger's equation:
\begin{equation}
E \begin{pmatrix}
		\Psi_{A}(\vec{k})\\
		\Psi_{B}(\vec{k})
  \end{pmatrix} = 	
\begin{pmatrix}
0 & f  \\
f^* & 0
	\end{pmatrix}
	\begin{pmatrix}
						\Psi_{A}(\vec{k}) \\
						\Psi_{B}(\vec{k}) \\
			\end{pmatrix}
\end{equation}

In a strained graphene, the distance vectors are modified by uniaxial strain as
$\vec{d}^s_l = (I+\epsilon) \vec{d}_l$ \cite{pereira2009tight}, where $l=1, 2, 3$. They are given by
\begin{align}
&\vec{d}^s_1 = -a \epsilon_{12} \hat{x} - (1+\epsilon_{22})a \hat{y}  \notag \\
&\vec{d}^s_2 = \frac{a}{2} \left[ (1+\epsilon_{11}) \sqrt{3} + \epsilon_{12} \right] \hat{x} +
	\frac{a}{2} \left[ \sqrt{3} \epsilon_{21} + (1+\epsilon_{22}) \right]\hat{y}   \notag \\
&\vec{d}^s_3 = \frac{a}{2} \left[ -(1+\epsilon_{11}) \sqrt{3} + \epsilon_{12} \right] \hat{x} +
	\frac{a}{2} \left[ -\sqrt{3} \epsilon_{21} + (1+\epsilon_{22}) \right]\hat{y} .
	\label{ds3}
\end{align}
Besides, the three hopping parameters $t_l$ are also altered by the strain, as discussed in the main text. They are determined by $t_l = t e^{-3.37(d_l^{s}/a-1)}$.
Then the geometric factor $f_s$ of the strained graphene is altered which can be calculated by $f_s = \sum_l^3  t_l \ e^{i \vec{k} \cdot \vec{d}_l^s}$. Taking into account the strain dressed hopping parameters and form factor, the correspondent equations of motion of an electron in the strained graphene can be obtained through substituting $f$ in Eq. (\ref{final_esq_1}) by $f_s$.
\begin{align}
f_s = & - t_1 e^{ \left[- i \epsilon_{12}k_x a - i (1+\epsilon_{22})k_y a\right]} \nonumber \\
	- & t_2 e^{i \left\{ \left[ \sqrt{3}(1+\epsilon_{11}) + \epsilon_{12} \right]k_x  + \left[(1+\epsilon_{22}) + \sqrt{3} \epsilon_{21} \right]k_y\right\}\frac{a}{2}}  \nonumber \\
	-& t_3 e^{i \left\{ \left[ -\sqrt{3}(1+\epsilon_{11}) + \epsilon_{12} \right]k_x  + \left[(1+\epsilon_{22}) - \sqrt{3} \epsilon_{21} \right]k_y\right\}\frac{a}{2}}
\end{align}

\subsection{Strained graphene with intrinsic SOC}
The pristine graphene with intrinsic SOCs can be well described by the following Hamiltonian:
\begin{align}
H = H_0 + H_{so}.
\label{hamilso}
\end{align}
The second term in the Eq. (\ref{hamilso}) is intrinsic SOC Hamiltonian. It can be evaluated by a summation over the next nearest-neighbors, as follows,
\begin{align}
H_{so} = \frac{2i}{\sqrt{3}} \lambda_{so} \sum_{\left< \left< i,j \right> \right>} \sum_{ \sigma, \sigma'} \ket{A_i, \sigma} \vec{\gamma} \cdot (\vec{d}_{kj} \times \vec{d}_{ik}) \bra{A_j, \sigma'} + h.c.,
\label{ham_ios}
\end{align}
where $\lambda_{so}$ is intrinsic SOC strength, $<<  >>$ runs over all the next-nearest neighbor sites as shown in Fig. \ref{fig7}(b) and $\vec{d}_{kj} \times \vec{d}_{ik}$ is a product of interatomic distances which represents an electron hopping from a $i$-site to a $j$-site through a $k$-site atom, as shown in Fig.\ref{fig7}(c). Performing the summation and other algebra calculations, one derives the equations of motion as,
\begin{align}
E \Psi_{A}(\vec{k},\sigma)= f \cdot \Psi_{B}(\vec{k},\sigma) - \lambda_{so} \cdot f_{so} \cdot \chi \Psi_{A}(\vec{k},\sigma), \notag
\end{align}
and
\begin{align}
E \Psi_{B}(\vec{k},\sigma)= f^* \cdot \Psi_{A}(\vec{k},\sigma) + \lambda_{so} \cdot f_{so} \cdot \chi \Psi_{B}(\vec{k},\sigma)
\end{align}
where
\begin{align}
f_{so} = 4 \sin\left(\frac{k_x \sqrt{3}}{2}a \right)  \nonumber
				&\left\{ cos\left(k_y \frac{3}{2} a \right) - cos\left(\frac{k_x \sqrt{3}}{2}a \right)\right\}
\end{align}
and $\chi = + 1$, for $\sigma = \uparrow$ or $\chi = - 1$, for $\sigma = \downarrow$. One can straightforwardly derive the matrix form of the Schr\"odinger's equation:
\begin{equation}
E \begin{pmatrix}
\Psi_{A}(\vec{k},\uparrow)\\
\Psi_{A}(\vec{k},\downarrow)\\
\Psi_{B}(\vec{k},\uparrow)\\
\Psi_{B}(\vec{k},\downarrow)
		\end{pmatrix}=	\begin{pmatrix}
- \varphi_{so} & 0 & f & 0 \\
0 &  \varphi_{so} & 0 & f & \\
f^* & 0 & \varphi_{so} & 0 \\
0 & f^* & 0  & -\varphi_{so}
		\end{pmatrix}
 \begin{pmatrix}
\Psi_{A}(\vec{k},\uparrow)\\
\Psi_{A}(\vec{k},\downarrow)\\
\Psi_{B}(\vec{k},\uparrow)\\
\Psi_{B}(\vec{k},\downarrow)
		\end{pmatrix}
\end{equation}	
where $\varphi_{so}=\lambda_{so} \cdot f_{so}$.

In strained graphene, since $C$-$C$ atomic distances are altered by applied strain, so does the product $\vec{d}_{kj} \times \vec{d}_{ik}$.  Taking into account the strain $dressed$ interatomic distances and  hopping parameters, the equations of motion of an electron in the strained graphene with intrinsic SOC become
\begin{align}
E \Psi_{A}(\vec{k},\sigma) = f_s \cdot \Psi_{B}(\vec{k},\sigma) - \lambda_{so}\cdot f_{sso} \cdot \chi\Psi_{A}(\vec{k},\sigma) \notag
\end{align}
and
\begin{align}
E \Psi_{B}(\vec{k},\sigma) = f^*_s \cdot \Psi_{A}(\vec{k},\sigma) + \lambda_{so}\cdot f_{sso} \cdot \chi\Psi_{B}(\vec{k},\sigma)
\end{align}
where the form factor of the strained graphene with intrinsic SOC is governed by
\begin{align}
f_{sso} = i\cdot det(I + \epsilon) \sum_{l=1}^3 \xi(l) \left( e^{i \vec{k} \cdot \vec{d}_l } - e^{-i \vec{k} \cdot \vec{d}_l } \right)   ,
\end{align}
with the pseudo-spin $\xi(l) = +1$, when $l=1, 3$ and $\xi(l) = -1$ for $l=2$.

\subsection{Strained graphene with Rashba SOC}
The breakdown of mirror symmetry induces the Rashba spin-orbit coupling. The Hamiltonian of a pristine graphene with Rashba SOC is well described by,
\begin{align}
	H = H_0 + H_{R}  .
\label{hamilr}
\end{align}
The second term in the Eq. (\ref{hamilr}) is Rashba SOC Hamiltonian. It can be evaluated by the following expression
\begin{align}
 H_R = i \sum_{\left< i,j \right>} \sum_{\sigma \sigma'} \left[ \ket{A_{i},\sigma} (\vec{u}_{ij} \cdot \vec{\gamma}) \bra{B_{j},\sigma'} + h.c. \right]   ,
\label{hamRashba}
\end{align}
where $\vec{u}_{ij} = - \frac{\lambda_R}{a} \hat{z} \times \vec{d}_{ij}$ and $\hat{z}$ is unit vector along z-axis. Then the correspondent equations of motion turn out to be:
\begin{align}
E \Psi_{A}(\vec{k},\uparrow) = f \cdot \Psi_{B}(\vec{k},\uparrow) - \lambda_R \cdot f_{R1} \cdot \Psi_{B}(\vec{k},\downarrow) \notag
\end{align}
and
\begin{align}
E \Psi_{B}(\vec{k},\uparrow) = f^* \cdot \Psi_{A}(\vec{k},\uparrow) + \lambda_R \cdot f_{R2} \cdot \Psi_{A}(\vec{k},\downarrow)
\end{align}
where the form factors of the graphene with Rashba SOC are defined by
\begin{align}
f_{R1} = i \left\{ e^{- i k_y 3a/2}
	+ 2 cos \left( k_x \frac{a_0}{2} - \frac{2\pi}{3} \right)  \right\} e^{i k_y \frac{a}{2}} \\
f_{R2} = i \left\{ e^{ i k_y 3a/2}
	+ 2 cos \left( k_x \frac{a_0}{2} + \frac{2\pi}{3} \right)  \right\} e^{-i k_y \frac{a}{2}}.
\end{align}
In analogy, we can deduce another set of coupled equations for $\Psi_{A}(\vec{k},\downarrow)$ and $\Psi_{B}(\vec{k},\downarrow)$.  After that, the matrix form of the Schr\"odinger's equation can be written as
\begin{equation}
E \begin{pmatrix}
\Psi_{A}(\vec{k},\uparrow)\\
\Psi_{A}(\vec{k},\downarrow)\\
\Psi_{B}(\vec{k},\uparrow)\\
\Psi_{B}(\vec{k},\downarrow)
		\end{pmatrix}=	\begin{pmatrix}
0 & 0 & f & \varphi_{R1} \\
0 & 0 & \varphi_{R2}^* & f & \\
f^* & \varphi_{R2} & 0 & 0 \\
\varphi_{R1}^* & f^* & 0 & 0
		\end{pmatrix}
 \begin{pmatrix}
\Psi_{A}(\vec{k},\uparrow)\\
\Psi_{A}(\vec{k},\downarrow)\\
\Psi_{B}(\vec{k},\uparrow)\\
\Psi_{B}(\vec{k},\downarrow)
		\end{pmatrix}
\end{equation}	
where was made $\varphi_{R1}=\lambda_{R} \cdot f_{R1}$ and $\varphi_{R2}=\lambda_{R2} \cdot f_{R2}$.

In the strained graphene with Rashba SOC, the corresponding Schr\"odinger's equation can be obtained by making the following substitutions: $f \rightarrow f_s$, $f_{R1} \rightarrow f_{sR1}$ and $f_{R2} \rightarrow f_{sR2}$. Here the form factors of the strained graphene with Rashba SOC are defined by
\begin{align}
f_{sR1} =  & - \frac{i}{a} \cdot \sum_{l=1}^3 \left[ e^{ i \vec{k} \cdot \vec{d}^s_l} \left( d^s_{ly} + i d^s_{lx} \right) \right],
\notag \\
f_{sR2} =  & \frac{i}{a} \cdot \sum_{l=1}^3 \left[ e^{ i \vec{k} \cdot \vec{d}^s_l} \left( d^s_{ly} + i d^s_{lx} \right) \right],
\end{align}
with $\vec{d}^s_{li}$ being the component of $\vec{d}^s_l$ vector along the $i$- direction with $i=x, y$.

\vspace{0.5cm}

\subsection{Strained graphene with exchange field}

\indent

Considering the pristine (unstrained) graphene subjected to an  exchange field, one may write
\begin{align}
	H = H_0 + H_{M}.
\label{hamilex}
\end{align}
The second term in the Eq. (\ref{hamilex}) is exchange Hamiltonian which is described by the following expression,
\begin{align}
H_M = M \sum_{i}^{N} \left\{ \left|A_i, \sigma \right> \gamma_{z} \left<A_i, \sigma \right| +h.c. \right\}
\end{align}
with strength $M$. The equations of motion is given by
\begin{align}
E & \Psi_{A}(\vec{k}, \sigma)  =  f \cdot \Psi_{B}(\vec{k},\sigma)+M \sigma \Psi_{A}(\vec{k},\sigma), \nonumber \\
E & \Psi_{B}(\vec{k}, \sigma)  =  f^{*} \cdot \Psi_{A}(\vec{k},\sigma)+M \sigma \Psi_{B}(\vec{k},\sigma).
\label{a_up}
\end{align}
We can also derive the matrix form of the Schr\"odinger's equation as
\begin{equation}
E \begin{pmatrix}
\Psi_{A}(\vec{k},\uparrow)\\
\Psi_{A}(\vec{k},\downarrow)\\
\Psi_{B}(\vec{k},\uparrow)\\
\Psi_{B}(\vec{k},\downarrow)
		\end{pmatrix}=	\begin{pmatrix}
M & 0 & f & 0 \\
0 & -M & 0 & f \\
f^* & 0 & M & f \\
0 & f^* & f^* & -M
		\end{pmatrix}
 \begin{pmatrix}
\Psi_{A}(\vec{k},\uparrow)\\
\Psi_{A}(\vec{k},\downarrow)\\
\Psi_{B}(\vec{k},\uparrow)\\
\Psi_{B}(\vec{k},\downarrow)
		\end{pmatrix}.
\end{equation}

In the strained graphene, the correspondent Schr\"odinger's equation can be obtained by doing a substitution of $f$ by $f_s$.

\subsection{Strained graphene with SOCs and exchange field}

With all effects together in the pristine (unstrained) graphene, the Hamiltonian now reads
\begin{align}
	H = H_0 + H_{so} + H_{R} + H_{M}   .
\label{hamilNanoZ}
\end{align}
Thus, the equation of motion is given by
\begin{align}
E \Psi_{A}(\vec{k},\sigma) = & f \cdot \Psi_{B}(\vec{k},\sigma)- \left[\lambda_{so}\cdot f_{so} - M \sigma\right] \cdot \Psi_{A}(\vec{k},\sigma)\nonumber \\
&-\lambda_R \cdot f_{R1} \cdot \Psi_{B}(\vec{k},-\sigma),
\end{align}
\begin{align}
E \Psi_{B}(\vec{k},\sigma) = & f^{*} \cdot \Psi_{A}(\vec{k},\sigma) + \left[\lambda_{so}\cdot f_{so} + M \sigma\right] \cdot \Psi_{B}(\vec{k},\sigma)\nonumber \\
&+\lambda_R \cdot f_{R2} \cdot \Psi_{A}(\vec{k},-\sigma).
\end{align}
The matrix form of the correspondent Hamiltonian with eigenvectors
\begin{equation}
(\Psi_{A}(\vec{k},\uparrow),\Psi_{A}(\vec{k},\downarrow),\Psi_{B}(\vec{k},\uparrow),\Psi_{B}(\vec{k},\downarrow))^{\dagger}
\end{equation}
reads
\begin{equation}
H =	\begin{pmatrix}
M-\varphi_{so} & 0 & f & \varphi_{R1} \\
0 & -M+\varphi_{so} & \varphi^*_{R2} & f \\
f^* & \varphi_{R2} & M+\varphi_{so} & f \\
\varphi^*_{R1} & f^* & f^* & -M-\varphi_{so}
		\end{pmatrix}.
\end{equation}

In the strained graphene, the correspondent Schr\"odinger's equation can be obtained by doing following substitutions: $f \rightarrow f_s$, $f_{so} \rightarrow f_{sso}$, $f_{R1} \rightarrow f_{sR1}$ and $f_{R2} \rightarrow f_{sR2}$.
\section{Equation of Motion of an Electron in Strained GNR}
In order to write the Hamiltonian for a GNR with zigzag edges, we must consider a unit cell $m$ and label each zigzag chain with parameter $n$. The whole Hamiltonian have the same general form of Eq. \ref{hamilNanoZ}. But the specific expression of each term in the Hamiltonian is different with its partner of the pristine graphene, given by
\begin{align}
	H_0 = & - \sum_{m,n}^{N} \sum_{\sigma,\sigma'}   \left\{ t_1 \left|A,m,n,\sigma \right> \left<B,m,n-1,\sigma' \right| + \right. \nonumber \\
		& + t_2 \left|A,m,n,\sigma \right> \left<B,m+1/2,n,\sigma' \right| + \nonumber \\
			& \left. + t_3 \left|A,m,n,\sigma \right> \left<B,m-1/2,n,\sigma' \right| + h.c \right\}.
\end{align}
As mentioned before, $t=2.7 eV$ for unstrained graphene. The intrinsic SOC is
\begin{align}
	H_{so}& = \nonumber \\
	& \frac{2i\lambda_{so} }{\sqrt{3}}  \sum_{m,n}^{N} \sum_{\sigma,\sigma'}  \left\{  \left|A_m,n,\sigma\right> \vec{\gamma} \cdot (\vec{d}^s_{kj} \times \vec{d}^s_{ik}) \left<A_{m+1},n,\sigma'\right| \right. \nonumber \\
	& \left|A_m,n,\sigma\right> \vec{\gamma} \cdot (\vec{d}^s_{kj} \times \vec{d}^s_{ik}) \left<A,m+1/2,n-1,\sigma'\right| + \nonumber \\
	& \left|A_m,n,\sigma\right> \vec{\gamma} \cdot (\vec{d}^s_{kj} \times \vec{d}^s_{ik}) \left<A_{m+1/2},n+1,\sigma'\right| + \nonumber \\
	& \left|A_m,n,\sigma\right> \vec{\gamma} \cdot (\vec{d}^s_{kj} \times \vec{d}^s_{ik}) \left<A_{m-1/2},n+1,\sigma'\right| + \nonumber \\
	& \left|A_m,n,\sigma\right> \vec{\gamma} \cdot (\vec{d}^s_{kj} \times \vec{d}^s_{ik}) \left<A_{m-1},n,\sigma'\right| + \nonumber \\
	& \left. \left|A_m,n,\sigma \right> \vec{\gamma} \cdot (\vec{d}^s_{kj} \times \vec{d}^s_{ik}) \left<A_{m-1/2},n-1,\sigma'\right| + h.c \right\}
\label{hamil_ISO}
\end{align}
with distance vectors modified by strain $\vec{d}^s_{kj}$, shown in Eq. \ref{ds3}. The Rashba SOC reads
\begin{align}
	H_R = & i \sum_{\left< m,n \right>}^{N} \sum_{\sigma \sigma'} \left[ \ket{A_m,n,\sigma} (\vec{u}_{n(n-1)} \cdot \vec{\gamma}) \bra{B_m,n-1,\sigma'} + \right. \nonumber \\
		& + \ket{A_m,n,\sigma} (\vec{u}_{m(m+1/2)} \cdot \vec{\gamma}) \bra{B_{m+1/2},n,\sigma'} +  \nonumber \\
		& + \left.  \ket{A_m,n,\sigma} (\vec{u}_{m(m-1/2)} \cdot \vec{\gamma}) \bra{B_{m-1/2},n,\sigma'} + h.c \right]
\end{align}
where $\vec{u}$ was modified to
\begin{align}
	&\vec{u}_{n(n-1)} = - \frac{\lambda_R}{a} \hat{z} \times \vec{d}^s_{n(n-1)}; \nonumber\\
	&\vec{u}_{m(m+1/2)} = - \frac{\lambda_R}{a} \hat{z} \times \vec{d}^s_{m(m+1/2)}; \\
	&\vec{u}_{m(m-1/2)} = - \frac{\lambda_R}{a} \hat{z} \times \vec{d^s}_{m(m-1/2)}.\nonumber
\end{align}
The EX term now is given by
\begin{align}
H_M = M  \sum_{m,n}^{N} \sum_{\sigma}\left\{ \left|A_m,n,\sigma \right> (\vec{\gamma} \cdot \hat{z}) \left<A_m,n,\sigma \right| +h.c. \right\}.
\end{align}

The wavevector now includes the periodicity of the unit cell
\begin{align}
	\left|\Psi\right> = \frac{1}{\sqrt{M}}  \sum_{m,n}^{N} \sum_{\sigma} e^{i \vec{k} \cdot \vec{R}_m} \left\{ \Psi_{A}(\vec{k},n, \sigma) \left|A_m,n, \sigma \right> + \right. \nonumber \\ \left. \beta (\vec{k},n, \sigma) \left|B_m,n, \sigma \right> \right\}.
	\label{zigzag_wavevector}
\end{align}
where $\vec{R}_m = m \vec{a}_0$ and $\vec{R}^{s}_m = (I+\epsilon) \vec{R}_m$ are the quantized distance among atoms in the absence and presence of strain, with $\vec{a}_0 = \sqrt{3}a \hat{x}$. Inserting this single-particle wavefunction into the Schr\"odinger equation, we obtain the following two equations of motion, 
\begin{align}
E & \Psi_{A}(\vec{k},n, \sigma)  =  - \left[   \beta (\vec{k},n, \sigma) \left( t_2 e^{i \omega} +  t_3 e^{-i \omega} \right) \right. \nonumber \\
    & + \left. t_1 \beta (\vec{k},n-1, \sigma)  \right] + 2 \chi \lambda_{so} det(I+ \epsilon)  \nonumber \\
	&  \left\{  \sin \left( \omega \right) \left[ \alpha(\vec{k},n-1,\sigma) + \alpha(\vec{k},n+1, \sigma) \right] \right. \nonumber \\
 &- \left. \sin \left( 2 \omega \right)  \alpha(\vec{k},n, \sigma) \right\}
	 	 - i \lambda_R \chi \left\{  \left[ - \rho_1 cos (\omega)  \right. \right. \nonumber \\
 	& + \left. \left. \sqrt{3} \rho_2 sen (\omega) \right] \beta(\vec{k},n,-\sigma)  + \rho_1 \beta(\vec{k},n,-\sigma) \right\} \nonumber \\
	& + \chi M \alpha(\vec{k},n,\sigma))
	\label{a_up}
\end{align}
and
\begin{align}
E & \Psi_{B}(\vec{k},n, \sigma)  = - \left[   \alpha (\vec{k},n,\sigma) \left( t_2 e^{i \omega} +  t_3 e^{-i \omega} \right) \right. \nonumber \\
    & + \left. t_1 \alpha (\vec{k},n+1,\sigma)  \right] - 2 \chi \lambda_{so} det(I+ \epsilon) \nonumber \\
	&  \left\{  \sin \left(\omega \right) \left[ \beta(\vec{k},n-1,\sigma) + \beta(\vec{k},n+1, \sigma) \right] \right. \nonumber \\
	& - \left. \sin \left(  2 \omega \right)  \beta(\vec{k},n, \sigma) \right\}
	+ i \lambda_R \chi \left\{ \left[ - \rho_1 cos (\omega)  \right. \right. \nonumber \\
	& - \left. \left. \sqrt{3} \rho_2 sen (\omega) \right] \alpha(\vec{k},n,-\sigma)  + \rho_1 \alpha(\vec{k},n,-\sigma) \right\} \nonumber \\
	& + \chi M \beta(\vec{k},n,\sigma))
	\label{b_up}
\end{align}
where
\begin{align}
\omega & = \frac{a_0}{2} \{ (1+ \epsilon_{11}) k_x + \epsilon_{21} ky \};\nonumber \\
\rho_1 &=  1 + \epsilon_{22} + i\epsilon_{12}; \nonumber \\
\rho_2 &= 1 + \epsilon_{11} - i \epsilon_{21}. \nonumber
\end{align}
%
\bibliography{references}
\end{document}